\documentclass[epjc3,twocolumn]{svjour3}          %

\usepackage{xcolor}
\usepackage{float}
\usepackage{array} 

\RequirePackage{eucal}

\RequirePackage{amsmath}
\RequirePackage{amssymb}
\RequirePackage[T1]{fontenc}

\RequirePackage{graphicx}
\RequirePackage{mathptmx}      
\RequirePackage{flushend}
\RequirePackage[numbers,sort&compress]{natbib}
\RequirePackage[colorlinks,citecolor=blue,urlcolor=blue,linkcolor=blue]{hyperref}

\journalname{Eur. Phys. J. C}
\usepackage{ulem}

\RequirePackage{xcolor}

\begin{document}

\def\nat{Nature}
\def\prl{Phys. Rev. Lett.}
\def\prb{Phys. Rev. B}
\def\prc{Phys. Rev. C}
\def\prd{Phys. Rev. D}

\def\mnras{Mon. Not. Roy. Astr. Soc.}
\def\apj{Astrophys. J.}
\def\apjl{Astrophys. J. Lett.}
\def\apjs{Astrophys. J. Suppl. Ser.}
\def\aa{Astron. Astrophys.}
\def\aap{Astron. Astrophys.}
\def\actaa{Acta Astronomica}
\def\aapr{Astron. Astrophys. Rev.}
\def\plb{Phys. Lett. B}
\def\pr{Phys. Rev.}
\def\araa{Annual Rev. of Astron. Astrophys.}

\def\pasj{Publications of the Astronomical Society of Japan }
\def\pasp{Publications of the Astronomical Society of the Pacific}
\def\zap{Zeitschrift f{\"u}r Astrophysik}

\def\npa{Nuclear Physics A}
\def\nphysa{Nucl. Phys.}
\def\physrep{Phys. Rep.}
\def\jcap{Journal of Cosmology and Astroparticle Physics}

\def\beq#1{\begin{equation}\label{#1}}
\def\eeq{\end{equation}}

\newcommand{\bear}[1]{\begin{eqnarray}\label{#1}}
\newcommand{\ear}{\end{eqnarray}}

\newcommand{\R}{{\mathbb R}}
\newcommand{\p}{\partial}
\newcommand{\nn}{\nonumber}

\title{Quasinormal modes in the field of a dyon-like dilatonic black hole \\ 
 corresponding to $A_1 + A_1$ Toda chain}

\author{ U. S.~ Kayumov\thanksref{e1,addr1} 
        \and
        S. V. ~Bolokhov\thanksref{e2,addr3}
        \and
        V. D.~Ivashchuk\thanksref{e3,addr2,addr3}
        \and
        A. N.~Malybayev\thanksref{e4,addr1} 
         \and \\
        G. S.~Nurbakova \thanksref{e5,addr1}
}
\thankstext{e1}{e-mail: ulqkayum@gmail.com}
\thankstext{e2}{e-mail: bol-rgs@yandex.ru}
\thankstext{e3}{e-mail: ivashchuk@mail.ru}
\thankstext{e4}{e-mail: algis\_malybayev@mail.ru}
\thankstext{e5}{e-mail: g.nurbakova@gmail.com}

\institute{ Department of Theoretical and Nuclear Physics, Al-Farabi Kazakh National University,  \\
Al-Farabi ave., 71, Almaty 050040, Kazakhstan\label{addr1}
\and
Center for Gravitation and Fundamental Metrology, \label{addr2}
\and
Institute of Gravitation and Cosmology, Peoples' Friendship University of Russia (RUDN University),
\\ 6 Miklukho-Maklaya St., Moscow 117198, Russian Federation\label{addr3}
}

\date{Received: date / Accepted: date}

\maketitle

\begin{abstract}
 We explore the quasinormal modes of a massless test scalar field in the gravitational background 
 of a non-extremal dilatonic dyonic black hole. The dyon-like black hole solution is considered 
 within a four-dimensional gravitational model that includes two scalar fields and two 2-forms. 
 It is governed by two dilatonic coupling vectors \(\vec{\lambda}_s\) (with \(s=1,2\)) 
 that satisfy \(\vec{\lambda}_1 = \vec{\lambda}_2 = \vec{\lambda}\) and \(\vec{\lambda}^2 = 1/2\).
 We derive and examine the quasinormal modes of a massless scalar test field in the eikonal limit. 
These modes are governed by the solution parameters $\mu > 0$, $P_1 > 0$, and $P_2 > 0$. 
In the two limiting scenarios -- namely $P_1 = P_2 = 0$ and $P_1 = P_2 = P > 0$ 
-- our results reduce exactly to the known (eikonal) quasinormal spectra for the Schwarzschild
 and Reissner-Nordström black holes, respectively.
Numerical higher-order WKB analysis of quasinormal frequencies for two lower levels (for various values of parameters) is also presented.  
\end{abstract}

\section{Introduction}

The recent detection of gravitational waves \cite{2016PhRvL.116f1102A,2019PhRvX...9c1040A,2020ApJ...892L...3A} has revived the longstanding interest in quasinormal modes (QNMs) \cite{1970Natur.227..936V,1971ApJ...170L.105P,1975RSPSA.344..441C,1984PhLA..100..231B,1984PhRvL..52.1361F,1984PhRvD..30..295F,1999LRR.....2....2K,1999CQGra..16R.159N,2009CQGra..26p3001B,2011RvMP...83..793K}, originally predicted by Vishveshwara in 1970. The observed signals were emitted during the ringdown phase of binary black hole mergers, and their frequencies are determined by superpositions of damped oscillations—namely, QNMs. A careful examination of these detections is of considerable importance, as it may provide crucial clues about the behavior of gravity in the strong-field regime.

In a nutshell, the (typical) quasinormal  modes (QNMs)  can  be obtained from the solutions to a wave equation  for  a scalar function  $\Phi(t,x)$ 
\begin{equation}
\label{0.0}
\Phi(t,x) = e^{-i\omega t} \Phi_{*} (x) 
 \end{equation}
with $\Phi_{*}   = \Phi_{*}  (x)$ obeying  a Schr\"odinger-type equation 
\begin{equation}
\label{0.1}
\left(- a^2\frac{d^2}{dx^2} + V(x)\right) \Phi_{*}  = \omega^2 \Phi_{*}
\end{equation}
on a (typical)  domain  $\mathbb{R}= (- \infty, + \infty)$, where $a > 0$ is  some parameter, e.g. $a = 1$, see Refs. \cite{1999LRR.....2....2K,1999CQGra..16R.159N,2009CQGra..26p3001B,2011RvMP...83..793K} and references there in. 

For  a certain class of asymptotically flat spherically symmetric   black-hole solutions 
(which contain Schwarzschild and Reissner-Nordstr\"om ones) the potential $V(x)$ is  positive with smooth dependence upon $x$. It has (typically) sufficiently fast fall off to zero when approaching either to the horizon: $x \to - \infty$ or to the spatial infinity: $x \to +\infty$. Here we denote by $x$ the so-called tortoise coordinate 
(in what follows it will be denoted as $R_{*}$).  
 
 The QNM cyclic frequencies $\omega$   obey ${\rm Re} \ \omega > 0$ and
${\rm Im} \ \omega < 0$, urging the wave functions (\ref{0.0})  to be exponentially damped in time 
as $t \to +\infty$ (describing the asymptotically  stable perturbations).  For  QNM  solutions 
to  equation (\ref{0.1}) the asymptotical conditions read as follows: the outgoing waves at spatial infinity
behave like $\Phi_{*}(x)   \sim e^{ \frac{i \omega x}{a}}$ for $x \to \infty$
  (${\rm  Re} \ \omega > 0$) while ingoing waves at the horizon obey 
 $\Phi_{*}(x)   \sim e^{- \frac{i\omega x}{a} }$ for $x \to-\infty$. 
 The absolute value $|\Phi_{*}(x)|$ is  exponentially growing (in $|x|$) as $|x| \to \infty$ due to restriction ${\rm Im} \ \omega < 0$.

   The standard analytical continuation method for QNMs \cite{2009CQGra..26p3001B,2011RvMP...83..793K,1984PhLA..100..231B,1984PhRvL..52.1361F,1984PhRvD..30..295F} has a transparent recent version in Ref. ~\cite{2020PhRvD.101b4008H}.  This method relates QNM frequencies of spherically symmetric black holes to bound-state energies of anharmonic oscillators via analytic continuation in $\hbar$. As shown in Ref.~\cite{2020PhRvD.101b4008H}, it reproduces WKB results, and the perturbative WKB series for the frequencies are   Borel-summable divergent series for both Schwarzschild and Reissner-Nordström black holes.
 
  This paper extends our previous studies \cite{ABDI,ABI,AIMT} 
    on dilatonic dyon/dyon-like black holes -- a subject also addressed in Refs.
 \cite{BronShikin,Gibbons,GM,GHS,ChHsuL,GKLTT,PTW,GKO,Dav,GalZad,MBI,IKMN}
  and references therein, originally motivated by (super)string/supergravity. We examine the QNM spectrum of a special non-composite 
  ($A_1 + A_1$) dyon-like solution from Ref. ~\cite{ABI}  (for composite case see Refs. \cite{ChHsuL,IKMN}) 
   in $4D$ with metric $g$, two scalars
  $\varphi^1,\varphi^2$, and two 2-forms $F^{(1)},F^{(2)}$ 
  carrying electric/magnetic charges $Q_1,Q_2$, with identical dilatonic coupling vectors $\vec{\lambda}_1=\vec{\lambda}_2=\vec{\lambda}$, $\vec{\lambda}^2=1/2$.
   Eikonal QNMs depend on  parameters $\mu>0, P_1>0, P_2>0$. 
   (QNMs of dilatonic holes are widely discussed in 
   Refs. \cite{Konoplya:2001ji,2001PhRvD..63f4009F,2002PhRvD..66h4007K,
   2005CQGra..22.1129C,2005JPhCS..24..123N,2013MPLA...2850109S,2015PhRvD..92f4022K,2018PhRvD..98j4042P,2019EPJC...79.1021B}.) 
   The 2-forms relate to charges by $F^{(1)}=Q_1\tau_1$, $F^{(2)}=Q_2\tau_2$, with $\tau_1$  and $\tau_2=\mathrm{vol}[S^2]$
   being electric and  magnetic $2$-forms, respectively. 
   For a single scalar with couplings $\lambda_1,\lambda_2$, the dyon-like ansatz was  treated in  Refs. \cite{ABI,AIMT,Dav,GalZad};  for $\lambda_1=\lambda_2=\lambda$, the  solutions from \cite{ABI} are  trivial non-composite generalizations of the one-2-form/one-scalar dyon solutions from Refs. \cite{ABDI,GKLTT,PTW,GKO}.

The paper is organised as follows: in section \ref{blackhole} we review the main properties of the black hole dyon-like solution  under consideration (a non-composite version of the solution from Ref. ~\cite{IKMN}). 
 In section \ref{parameters} we consider the physical parameters 
 and particular cases of the dyonic black hole solutions.  In section \ref{qnms} we deal with 
 QNM frequencies for massless test scalar field ``living''
 in the background metric of our dyon-like black hole solution. 
 Here we combine numerical WKB analysis of quasinormal frequencies at lower levels (subsection 4.1) 
 with analytical derivation of  the eikonal approximation (subsection 4.2) and
  two limiting cases (subsection \ref{limitingcase2}) $: P_1 = P_2= +0$ and $P_1 = P_2 = P > 0$, 
 which are associated with the Schwarzschild and Reissner-Nordstr\"om black hole solutions
 are also presented. A brief summary of our findings is then presented in section \ref{conclusion}.

\section{Black hole dyon solutions}\label{blackhole}

We start with the action of a model containing two scalar fields, two 2-forms and 
two dilatonic coupling vectors which reads as follows 
~\cite{MBI},  
\bear{i.1}
 S= \frac{1}{16 \pi G}  \int d^4 x \sqrt{|g|}\biggl\{ {\cal R}[g] -
  g^{\mu \nu} \p_{\mu} \vec{\varphi}  \p_{\nu} \vec{\varphi}
 \qquad \qquad   \nn \\
 - \frac{1}{2} e^{2 \vec{\lambda}_1 \vec{\varphi}} F^{(1)}_{\mu \nu} F^{(1)\mu \nu }
 - \frac{1}{2} e^{2 \vec{\lambda}_2 \vec{\varphi}} F^{(2)}_{\mu \nu} F^{(2) \mu \nu}
 \biggr\},
\ear
where $g= g_{\mu \nu}(x)dx^{\mu} \otimes dx^{\nu}$ is the metric,  $|g| =   |\det (g_{\mu \nu})|$, 
 $\vec{\varphi} =  (\varphi^1,\varphi^2)$ is the vector of scalar fields 
 belonging to ${\R}^2$, 
 $F^{(s)} = dA^{(s)}  =  \frac{1}{2} F^{(s)}_{\mu \nu} dx^{\mu} \wedge dx^{\nu}$
is the $2$-form with $A^{(s)} = A^{(s)}_{\mu} dx^{\mu}$, $s =1,2$; 
$G$ is the gravitational constant,
 $\vec{\lambda}_1 = (\lambda_{1i})$, 
 $\vec{\lambda}_2 = (\lambda_{2i})$ are the dilatonic coupling  vectors  
 and ${\cal R}[g]$ is the Ricci scalar.
    
 We adopt the following choice:
\beq{i.1a}
  \qquad \vec{\lambda_1} = \vec{\lambda_2} =  \vec{\lambda}, \quad \vec{\lambda}^2 = \frac{1}{2}.
\eeq
Additionally, we set $c = 1$, with $c$ being the speed of light in vacuum.

We consider  dyon black hole
solution to the field equations corresponding to the action
(\ref{i.1})  which is defined on the manifold
\begin{equation}  \label{i.2}
 {\cal M }  =    (2\mu, + \infty)  \times S^2 \times  \R,
\end{equation}
and has the following form \cite{ABI} 
\begin{eqnarray}  \nonumber
 ds^2 = g_{\mu \nu} dx^{\mu} dx^{\nu} \qquad \qquad
 \\ \nonumber
 = H_1 H_2
 \biggl\{ -  H_1^{-2 } H_2^{-2 } 
 \left( 1 - \frac{2\mu}{R} \right)
 dt^2 \\ 
  \qquad +  \frac{dR^2}{1 - \frac{2\mu}{R}} + R^2  d \Omega^2_{2}
  \biggr\},  \label{i.3}
 \\  \label{i.3a}
 \vec{\varphi} = \vec{\lambda} \ln (H_1/H_2),
 \\  \label{i.3bem}
 F^{(1)}=  \frac{Q_1}{R^2}   H_{1}^{-2}  dt \wedge dR, \qquad    F^{(2)} =  Q_2 \tau.
\end{eqnarray}

Here  $Q_1$ is electric charge and $Q_2$ is magnetic charge, 
$\mu > 0$,  $d \Omega^2_{2} = d \theta^2 + \sin^2 \theta d \phi^2$
is the  metric on the unit sphere $S^2$
 ($0< \theta < \pi$, $0< \phi < 2 \pi$),
 $\tau = \sin \theta d \theta \wedge d \phi$
is the  volume form on $S^2$. 
For the speed of light we put $c=1$.

The functions $H_s$ are given by
\begin{equation} \label{i4.2}
 H_s = H_s(R)  = 1 + \frac{P_s}{R},
\end{equation}
where
\begin{equation} \label{i4.3}
 P_s (P_s + 2 \mu) = Q_s^2,
\end{equation}
$s = 1,2$. By utilizing the  positive roots
from Eq. (\ref{i4.3})
\begin{equation} \label{i4.3p}
    P_s  = P_{s,+} =  - \mu + \sqrt{\mu^2 +  Q^2_s} > 0,
\end{equation}
we arrive at a well-defined solution  for $R > 2\mu$.

The moduli functions $H_s$ satisfy 
the following boundary conditions:
\begin{equation}  \label{i3.1a}
  H_s  \to H_{s0} = 1 + \frac{P_s}{2 \mu}  > 0
\end{equation}
for $R \to 2\mu $, and
\begin{equation} \label{i3.1b}
  H_s    \to 1
\end{equation}
for $R \to +\infty$, $s = 1,2$.

In brane terminology, this is a composite solution that describes a configuration 
of two intersecting non-extremal black $0$-branes -- an electric one 
and a magnetic one -- with the intersection rule corresponding 
to the Lie algebra $A_1 + A_1$ ($A_1 = sl(2)$) \cite{IMtop}.

According to the first boundary condition (\ref{i3.1a}), we obtain a (regular) horizon at 
$R = R_g = 2 \mu$ for the metric (\ref{i.3}). The second condition 
(\ref{i3.1b}) guarantees asymptotic flatness of the metric as 
$R \to +\infty$.

 Globally extending the metric (originally for $R>2\mu$) exposes horizons 
 at $R=2\mu$ and $R=0$, and a singularity at $R=-\min(P_1,P_2)$.
          
{\bf Symmetric $P_1 = P_2$ case.}
Let us put $P_1 = P_2 = P > 0$.
  We obtain
\beq{i4.11}
  Q_1^2 =  Q_2^2 = \frac{1}{2} Q^2 = P (P + 2 \mu). 
 \eeq
By using new radial variable, $R + P = r$, we obtain
 \bear{i4.12}
  &&ds^2 =   - f(r) dt^2 +   (f(r))^{-1} dr^2 + r^2  d \Omega^2_{2},
     \\    \label{i4.12FP}
  &&F^{(1)}= \frac{Q_1}{r^2} dt \wedge dr, \quad F^{(2)}  =  Q_2 \tau,
  \qquad   \vec{\varphi} = \vec{0},
    \ear
 where $f(r) = 1 - \frac{2GM}{r} + \frac{Q^2}{2r^2}$ and $GM = P + \mu$ =
 $\sqrt{\mu^2 + \frac{1}{2} Q^2}$. Thus, for $P_1 = P_2$   we are led to  the  Reissner-Nordstr\"om  
 metric governed by two  parameters: $GM > 0$ and $ Q_{RN}^2 = Q^2/2 <  (GM)^2$. 

\section{Physical parameters and black hole thermodynamics}\label{parameters}

Here, we provide specific physical parameters associated with the black hole solution.

\subsection{Gravitational mass and scalar charges}

  From Eq.~\eqref{i.3}, one derives the ADM gravitational mass in the weak field regime 
 by matching it to $g_{00} = - (1 - 2GM/R + O(1/R))$:
 \beq{i5.1}
  GM =   \mu + \frac{1}{2}(P_1 + P_2).
 \eeq

 The scalar charge vector $\vec{Q}_{\varphi} = ( Q_{\varphi}^1, Q_{\varphi}^2 )$, 
 in turn, is extracted from (\ref{i.3a}) in the weak field limit on the basis of the relation 
 $\varphi^i = Q_{\varphi}^i /R + O(1/R)$:
 \color{black}
 \beq{i5.1s}
  \vec{Q}_{\varphi} =    \vec{\lambda} (P_1 - P_2).
  \eeq
 
 Combining (\ref{i5.1}) and (\ref{i5.1s}) gives the following identity
 \begin{equation}
 \label{i5.1id}
      2 (GM)^2   +    \vec{Q}_{\varphi}^2   = Q_1^2 + Q_2^2 + 2 \mu^2.
 \end{equation}
 
 The above expression does not involve the vectors $\vec{\lambda}_s$.
 In the extremal limit $\mu = +0$, this relation was earlier reported in Ref. \cite{PTW}.
 
 {\bf Bounds on mass}. It may be readily verified that the following bounds on 
 BH mass take place (see also Ref. \cite{ABDI})
  \begin{equation}
  \label{5.1id}
  \frac{1}{4} (Q_1^2 + Q_2^2) < (GM)^2 \leq \frac{1}{2} (Q_1^2 + Q_2^2) + \mu^2.
  \end{equation}
    For  $P_1 = P_2 = P$ case we have a special (more subtle) bound  
   $Q_1^2 + Q_2^2 < 2 (GM)^2$ instead of the first inequality in Ref. (\ref{5.1id})

\subsection{Black hole thermodynamics}

In this subsection, we address black hole thermodynamics by computing the Hawking temperature and entropy, verifying the first law, and testing the Smarr relation.

Here   we put (for simplicity) $\hbar = c = k_B = 1$.
The Bekenstein–Hawking (area) entropy, $S = A/(4G)$, where $A$ is the horizon area, is associated with the black hole solution (\ref{i.3}) at the horizon $R = 2\mu$, and is given by 
\beq{i5.2s}
 S = S_{BH} =   \frac{4 \pi \mu^2}{G}  \left(1 + \frac{P_1}{2 \mu}\right) \left(1 + \frac{P_2}{2 \mu}\right),
\eeq
while the corresponding Hawking temperature is given by
\beq{i5.2}
  T = T_H =  \frac{1}{8 \pi \mu} \left(1 + \frac{P_1}{2 \mu}\right)^{-1} \left(1 + \frac{P_2}{2 \mu}\right)^{-1}.
\eeq
Due to (\ref{i5.2}) and (\ref{i5.2s}) we have 
 $T_H  S_{BH} =   \frac{\mu}{2G}$.

One can readily verify that relations (\ref{i5.1}), (\ref{i5.2s}), and (\ref{i5.2})
 imply the first law of black hole thermodynamics.
\beq{i5.4t}
   dM = T dS + \Phi_1 dQ_1 + \Phi_2 dQ_2,
\eeq
and the Smarr relation
\beq{i5.4S}
   M = 2 T S + \Phi_1 Q_1 + \Phi_2 Q_2,
\eeq
where 
\beq{i5.4P}
    \Phi_s =   \frac{ Q_s}{2 G (P_s + 2 \mu) },
\eeq
$s = 1,2$. 

We now clarify the physical interpretation of the potentials given in (\ref{i5.4P}). 
From the first relation in (\ref{i.3bem}) for \(F^{(1)} = dA^{(1)}\), 
a particular solution for the $1$-form is
\beq{i5.5A1}
A^{(1)} = A^{(1)}_0(R)\, dt = \frac{Q_1}{R + P_1}\, dt,
\eeq
which leads to
\beq{i5.5AP1}
\Phi_1 = \frac{1}{2G}\, A^{(1)}_0(2\mu).
\eeq
Thus, $\Phi_1$ is equal, up to the factor $1/(2G)$, to the horizon value of the time component 
of the first Abelian gauge field $A^{(1)}$ -- that is, the electric potential 
(in the chosen gauge) corresponding to the electric charge.

Now we turn to the magnetic term in (\ref{i.3bem}). Computing the Hodge dual gives
\beq{i5.6}
  *F^{(2)} = \frac{Q_2}{H_1 H_2 R^2} \, dt \wedge dR,
\eeq
where we adopt the convention \(*F_{\mu \nu} 
= \frac{1}{2} \sqrt{|g|} \varepsilon_{\mu \nu \rho \sigma} F^{\rho \sigma}\) 
with \(\varepsilon_{0123}=1\). From relation (\ref{i.3a}) we obtain
\beq{i5.8}
 e^{2\vec{\lambda}_2 \vec{\varphi}} = \frac{H_1}{H_2}.
\eeq
For the \(S\)-dual $2$-form defined as
\beq{i5.9}
 \Tilde{F}^{(2)} = d\Tilde{A}^{(2)} = e^{2\vec{\lambda}_2 \vec{\varphi}} \, *F^{(2)},
\eeq
we find
\beq{i5.10F}
 \Tilde{F}^{(2)} = \frac{Q_2}{H_2^{2} R^2} \, dt \wedge dR = \frac{Q_2}{(R+P_2)^2} \, dt \wedge dR.
\eeq
Accordingly, we may choose the corresponding $1$-form as
\beq{i5.10A}
 \Tilde{A}^{(2)} = \Tilde{A}^{(2)}_0(R) \, dt = \frac{Q_2}{R+P_2} \, dt.
\eeq
Hence,
\beq{i5.11}
  \Phi_2 = \frac{1}{2G} \, \Tilde{A}^{(2)}_0(2\mu),
\eeq
which means that $\Phi_2$ coincides, up to the factor $1/(2G)$, 
with the horizon value of the time component of the dual Abelian gauge field \(\Tilde{A}^{(2)}\)
-- that is, the dual electric potential (in the chosen gauge) -- which corresponds to the magnetic charge field modulated by the scalar fields.

\section{Quasinormal modes}\label{qnms}

In this section, we turn to the quasinormal modes of our static, spherically symmetric solution, 
whose metric may be  written in the general parametrization of the radial variable
\begin{equation}\label{QNM2m}
ds^2 = -A(u) dt^2 + B(u) du^2 + C(u) d\Omega^2,
\end{equation}
with $A(u), B(u), C(u)>0$ and $d\Omega^2 = d\theta^2 + \sin^2\theta \, d\phi^2$. 
Throughout this section and the remainder of the paper, we adopt Planck units (\(\hbar = G = c =1\)).

We consider a test massless scalar field propagating in the background described by the metric (\ref{QNM2m}). Its equation of motion is given by the covariant Klein–Fock–Gordon equation
\begin{equation}\label{QNM2}
\Delta \Psi \equiv \frac{1}{\sqrt{|g|}} \, \partial_\mu \left( \sqrt{|g|} \, g^{\mu\nu} \partial_\nu \Psi \right) = 0,
\end{equation}
where the indices $\mu,\nu$ run over $0,1,2,3$.

To solve this equation, we separate variables by writing
\begin{equation}\label{QNM3}
\Psi = e^{-i\omega t} e^{-\gamma} \Psi_*(u) Y_{lm},
\end{equation}
where \(Y_{lm}\) are the spherical harmonics. 
Substituting \eqref{QNM3} into \eqref{QNM2} yields the radial equation for \(\Psi_*(u)\), 
which takes a Schrödinger-like form:
\begin{equation}\label{QNM4}
\frac{d^2 \Psi_*(u)}{du^2} + \left\{ \frac{B}{A} \omega^2 - 
\frac{B}{C} l(l+1) - \gamma'' - (\gamma')^2 \right\} \Psi_*(u) = 0,
\end{equation}
where
\begin{equation} \label{gamma_1}
\gamma = \frac{1}{2} \ln\left(B^{-1} C \sqrt{AB}\right),
\end{equation}
and \(\gamma' = d\gamma/du\), with \(l\) being the multipole quantum number, \(l = 0,1,2,\dots\).

Using the above expressions, we can now investigate the dyon-like black hole solution, 
which has the following form
\begin{equation}
 ds^2 =-f(R) dt ^2 + \frac{dR ^2}{f(R)} + C(R) d \Omega ^2 \ , \\
\end{equation}
where $f(R)$ and $C(R)$ according to Eq.~\eqref{i.3}  can be written as
\begin{eqnarray}
 f(R)&=&A = (H_1 H_2)^{-1} \left( 1-\frac{2\mu}{R} \right)\ , \\
 C&=&H_1 H_2 R^2\ ,
\end{eqnarray}
where  $H_s(R)= 1 + P_s/R$, $s =1,2$,
are moduli functions, $\mu$, $P_1 > 0$ and $P_2 > 0$. 

After applying the tortoise coordinate transformation
\begin{equation} \label{tort}
dR_*=\frac{dR}{f}
\end{equation}
the metric becomes
\begin{equation} \label{ds2fC}
 ds^2 =-f dt ^2 + f dR_*^2 + C d \Omega ^2 \ . 
\end{equation}
By choosing the tortoise coordinate as the radial coordinate, 
i.e. $u = R_*$, we have $A = B = f$, and consequently
\begin{equation} \label{gamma_2}
\gamma =\frac{1}{2}\ln C=\frac{1}{2}\ln (H_1 H_2 R^2).
\end{equation}

Thus, the   radial equation \label{QNM4} becomes
\begin{equation}\label{QNM5}
\frac{d^{2}\Psi_{*}}{dR_*^{2}}+\big\{ \omega 
^{2}-V\big\} \Psi_{*}=0, 
\end{equation}
where $\omega$ is the cyclic frequency (of the quasinormal mode) and 
$V =V(R) = V(R(R_*))$ is the effective potential
 \begin{equation} \label{QNM5V}
 V= \mathcal{V}+\delta\mathcal{V},
 \end{equation}
where 
 \begin{equation} \label{QNM5Veik}
  \mathcal{V} =\frac{l(l+1)f}{H_1 H_2R^2}
 \end{equation}
 is  the eikonal part of the effective potential and
 \begin{eqnarray} 
  \delta\mathcal{V} = \gamma''+(\gamma ')^{2} 
=   \frac{R (R-2 \mu )}{4 (P_1+R)^4 (P_2+R)^4} \times \nonumber \\
 \times   [R(R (P_1^2+ 14 P_1 P_2+ P_2^2) +4 R^2 (P_1+ P_2)
 \nonumber \\
   +4 P_1 P_2 (P_1+ P_2))+2 \mu (R (P_1^2-6 P_1 P_2+ P_2^2)
  \nonumber \\
  +2 R^2 (P_1+P_2)-2 P_1 P_2 (P_1+ P_2)+
  4 R^3)]
 \label{QNM5Vneik}
 \end{eqnarray}
is non-eikonal part of the effective potential. 

Here and what follows we denote $F' = \frac{dF}{dR_{*}}= f \frac{dF}{dR}$.

\begin{figure*}[ht]
\centering
\begin{tabular}{lr}
\includegraphics[width=0.48\linewidth]{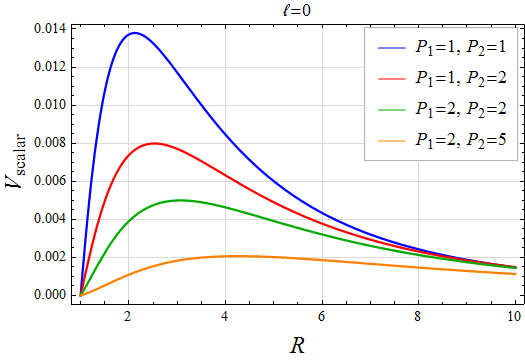} &\includegraphics[width=0.48\linewidth]{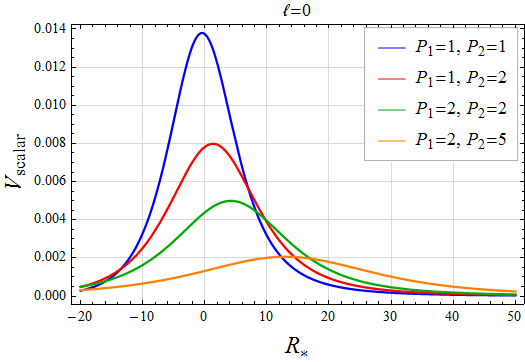}
\end{tabular}
\caption{Examples of plots of the effective potential \eqref{QNM5V} as a function of $R$ (left panel) and $R_*$ (right panel) for a particular set of the parameters $P_1, P_2$ and $\ell=0$, $\mu=1/2$.}

\label{Vplots}
\end{figure*}

\subsection{Numerical WKB analysis}

The quasinormal frequencies $\omega=\omega_\textbf{R}-i\omega_{\textbf{Im}}$ of test fields 
on a spherically-symmetric black hole background for different overtones $n$ and multipole numbers 
$\ell \gtrsim n$ can be effectively estimated using the so-called semi-analytical 
Wentzel--Kramers--Brillouin (WKB) approach. This method is mostly applicable when the effective potential
$V(R)$ exhibits a good single-peaked barrier reaching constant values at the boundaries $R_\star\to\pm\infty$
corresponding to the event horizon and spatial infinity. The WKB approach was first applied 
to the problem of finding quasinormal modes and grey-body factors of black holes by B. Schutz and C. Will.~\cite{schutz}.

For potential barriers characterized by a smooth profile and a unique maximum (which is typical for test fields in static, spherically symmetric backgrounds), the WKB technique offers a robust computational framework and is used in many works for calculating quasinormal spectra and grey-body factors for various field spins and black hole geometries (see~\cite{2009CQGra..26p3001B, 2011RvMP...83..793K, konJCAP2023, zinh, dubinsk, lutfu, malik, bolmil} and references therein).

The WKB method is based on matching the wave function expansions across three distinct regions 
(the vicinity of the potential peak and the two asymptotic sides) by ensuring the continuity of logarithmic derivatives at the turning points. In the eikonal limit ($\ell \to \infty$), where the field oscillates rapidly compared to the potential's spatial variation, the first-order (lowest) WKB method becomes exact. The formalism is extended to higher orders by treating the complex frequency as a systematic expansion in terms of the parameter $\mathcal{K}$, defined by the overtone number $n$ as
\begin{equation}
\mathcal{K}=n+\frac{1}{2}, \qquad n=0,1,2,\ldots
\end{equation}
By evaluating the potential and its higher-order derivatives $V_i=d^iV/dR_*^i$ at the peak point, the squared quasinormal
frequency is represented as an asymptotic series
\begin{align}
\omega^{2}
&= V_{0} + \Lambda_2(\mathcal{K}^2)+\Lambda_4(\mathcal{K}^2)+... \nonumber \\ 
& - i\mathcal{K}\sqrt{-2V_2}\left(1+\Lambda_3(\mathcal{K}^2)+\Lambda_5(\mathcal{K}^2)+...\right).
\end{align}
In this expression, the coefficients $A_i$ are cumbersome expressions involving the potential's derivatives up to order $2i$ at the peak point. Explicit analytical forms for these corrections have been developed in the literature up to the thirteenth WKB order ~\cite{Iyer, KonoplyaBH, Matyjasek}.

While the WKB expansion typically demonstrates rapid convergence for the fundamental mode ($n=0$) and low-order overtones
($n < \ell$), increasing the expansion order does not monotonically guarantee higher precision. To enhance numerical stability and accuracy, it is standard practice to apply Pad\'e resummation to the $\omega^2$ series~\cite{Matyjasek}. This Pad\'e-improved WKB method frequently yields results comparable in precision to sophisticated numerical schemes like Leaver’s (Frobenius's) continued-fraction method.

In this study, we employ the 6th- and 9th-order WKB formula with Pad\'e approximants to investigate the quasinormal spectrum of a test massless scalar field with the potential \eqref{QNM5V}. To check the reliability of the obtained numerical results, we present in Tables \ref{tableQNM0} and \ref{tableQNM1} both WKB6-Pad\'e and WKB9-Pad\'e schemes and estimate their relative difference in percentage. The results demonstrate excellent agreement between these two particular numerical WKB schemes, especially for the case $n=0$, $\ell=1$ (up to a small fraction of a percent).

\begin{figure*}[h!]
\centering
\begin{tabular}{lr}
\includegraphics[width=0.48\linewidth]{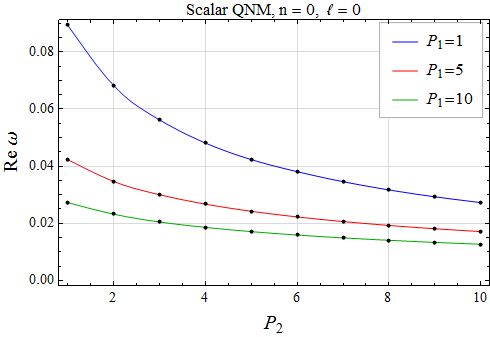} &\includegraphics[width=0.48\linewidth]{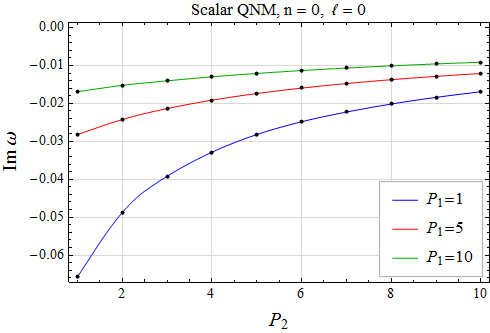}
\end{tabular}
\caption{Plots of the real part (Left panel) and the imaginary part (Right panel) of the fundamental  quasinormal mode ($n=0$) of the massless scalar field for a particular set of the parameters $P_1, P_2$ and $\ell=0$, $\mu=1/2$.}

\label{QNM0}
\end{figure*}

\begin{figure*}[h!]
\centering
\begin{tabular}{lr}
\includegraphics[width=0.48\linewidth]{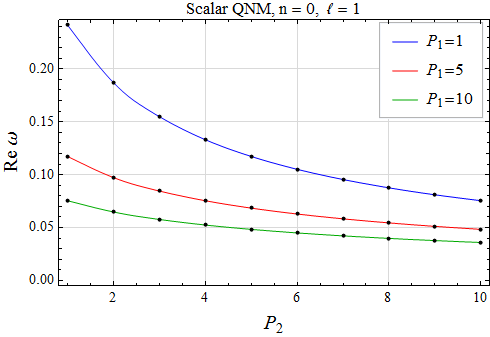} &\includegraphics[width=0.48\linewidth]{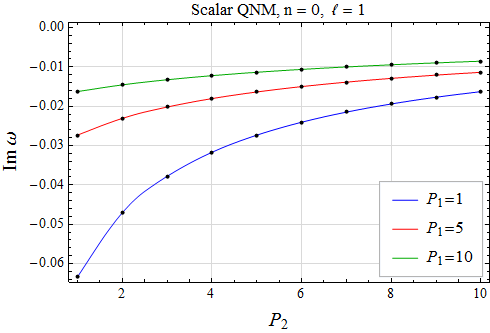}
\end{tabular}
\caption{Plots of the real part (Left panel) and the imaginary part (Right panel) of the fundamental quasinormal mode ($n=0$) of the massless scalar field for a particular set of the parameters $P_1, P_2$ and $\ell=1$, $\mu=1/2$.}

\label{QNM1}
\end{figure*}

The plots \ref{QNM0} and \ref{QNM1} illustrate the behaviour of the real and imaginary parts of the fundamental quasinormal frequency $\omega$ for $\ell=0$, $\ell=1$ and a particular set of the parameters $P_1, P_2$. One can see that, with increasing the parameters $P_1, P_2$, when the potential peak decreases, the oscillatory (real) part of the spectrum and the damping rate (imaginary part) have the tendency to decrease in absolute value. At the same time, in contrast to the oscillatory part, the imaginary part of the spectrum is less sensitive to changes in the multipole number $\ell$.

\begin{table}[ht]
\centering
\setlength{\tabcolsep}{2.5pt} 
\caption{Quasinormal frequencies of the scalar field with $n=0$, $\ell=0$, and various $P_1$, $P_2$.
(We put $ \mu = 1/2$.)  
Here $\tilde m$ denotes the order of the Pad\'e approximant. The results from the sixth- and ninth-order show good convergence with small differences $\sim$ below a percent.}

\begin{tabular}{lllll}
\hline
\text{P1} & \text{P2} & \text{WKBPade6($\tilde m=4$)} &
   \text{WKBPade9($\tilde m=4$)} & \text{Diff,$\%$} \\
   \hline
   \hline
 \text{      1} & \text{      1} & \text{ 0.0905151}-\text{ 0.0661586} i & \text{
   0.0893681}-\text{ 0.0657619} i & 1.09 \\
 \text{      1} & \text{      2} & \text{ 0.0690282}-\text{ 0.0482152} i & \text{
   0.0683198}-\text{ 0.0487703} i & 1.07 \\
 \text{      1} & \text{      3} & \text{ 0.0565201}-\text{ 0.0387071} i & \text{
   0.056206}-\text{ 0.0391777} i & 0.83 \\
 \text{      1} & \text{      4} & \text{ 0.0483734}-\text{ 0.0325408} i & \text{
   0.0481177}-\text{ 0.0328353} i & 0.67 \\
 \text{      1} & \text{      5} & \text{ 0.0425662}-\text{ 0.0281801} i & \text{
   0.0423068}-\text{ 0.0281909} i & 0.51 \\
 \text{      1} & \text{      6} & \text{ 0.0381044}-\text{ 0.0249379} i & \text{
   0.0380847}-\text{ 0.0248108} i & 0.28 \\
 \text{      1} & \text{      7} & \text{ 0.0345123}-\text{ 0.0223504} i & \text{
   0.0345777}-\text{ 0.0222483} i & 0.29 \\
 \text{      1} & \text{      8} & \text{ 0.0315957}-\text{ 0.0202105} i & \text{
   0.0316944}-\text{ 0.0201348} i & 0.33 \\
 \text{      1} & \text{      9} & \text{ 0.029187}-\text{ 0.0184291} i & \text{
   0.0292903}-\text{ 0.0183779} i & 0.33 \\
 \text{      1} & \text{     10} & \text{ 0.0271565}-\text{ 0.0169304} i & \text{
   0.027251}-\text{ 0.016894} i & 0.32 \\
 \text{      5} & \text{      1} & \text{ 0.0425662}-\text{ 0.0281801} i & \text{
   0.0423068}-\text{ 0.0281909} i & 0.51 \\
 \text{      5} & \text{      2} & \text{ 0.0348652}-\text{ 0.0241113} i & \text{
   0.0347023}-\text{ 0.0242429} i & 0.49 \\
 \text{      5} & \text{      3} & \text{ 0.0301838}-\text{ 0.0212423} i & \text{
   0.0300206}-\text{ 0.0213401} i & 0.52 \\
 \text{      5} & \text{      4} & \text{ 0.0268773}-\text{ 0.0190469} i & \text{
   0.026723}-\text{ 0.0191271} i & 0.53 \\
 \text{      5} & \text{      5} & \text{ 0.0243689}-\text{ 0.0172999} i & \text{
   0.0242264}-\text{ 0.0173714} i & 0.53 \\
 \text{      5} & \text{      6} & \text{ 0.0223783}-\text{ 0.0158701} i & \text{
   0.0222478}-\text{ 0.0159371} i & 0.54 \\
 \text{      5} & \text{      7} & \text{ 0.0207477}-\text{ 0.0146743} i & \text{
   0.0206285}-\text{ 0.0147389} i & 0.53 \\
 \text{      5} & \text{      8} & \text{ 0.0193804}-\text{ 0.0136569} i & \text{
   0.0192714}-\text{ 0.0137201} i & 0.53 \\
 \text{      5} & \text{      9} & \text{ 0.0182127}-\text{ 0.0127792} i & \text{
   0.0181128}-\text{ 0.0128412} i & 0.53 \\
 \text{      5} & \text{     10} & \text{ 0.0172009}-\text{ 0.0120133} i & \text{
   0.0171088}-\text{ 0.0120741} i & 0.53 \\
 \text{     10} & \text{      1} & \text{ 0.0271565}-\text{ 0.0169304} i & \text{
   0.027251}-\text{ 0.016894} i & 0.32 \\
 \text{     10} & \text{      2} & \text{ 0.0232147}-\text{ 0.0153195} i & \text{
   0.0232402}-\text{ 0.0152293} i & 0.34 \\
 \text{     10} & \text{      3} & \text{ 0.0205973}-\text{ 0.0139634} i & \text{
   0.0204833}-\text{ 0.0140189} i & 0.51 \\
 \text{     10} & \text{      4} & \text{ 0.0186835}-\text{ 0.0128986} i & \text{
   0.0185874}-\text{ 0.0129666} i & 0.52 \\
 \text{     10} & \text{      5} & \text{ 0.0172009}-\text{ 0.0120133} i & \text{
   0.0171088}-\text{ 0.0120741} i & 0.53 \\
 \text{     10} & \text{      6} & \text{ 0.0160008}-\text{ 0.0112579} i & \text{
   0.0159114}-\text{ 0.0113112} i & 0.53 \\
 \text{     10} & \text{      7} & \text{ 0.0150005}-\text{ 0.0106031} i & \text{
   0.0149141}-\text{ 0.0106505} i & 0.54 \\
 \text{     10} & \text{      8} & \text{ 0.0141486}-\text{ 0.0100283} i & \text{
   0.0140657}-\text{ 0.0100715} i & 0.54 \\
 \text{     10} & \text{      9} & \text{ 0.0134112}-\text{ 0.00951882} i & \text{
   0.0133318}-\text{ 0.00955898} i & 0.54 \\
 \text{     10} & \text{     10} & \text{ 0.0127643}-\text{ 0.00906331} i & \text{
   0.0126885}-\text{ 0.00910133} i & 0.54 \\
\hline
\hline
\end{tabular}
\label{tableQNM0}
\end{table}

\begin{table}[ht]
\centering
\setlength{\tabcolsep}{4pt}
\caption{Quasinormal frequencies of the scalar field with $n=0$, $\ell=1$, and various $P_1$, $P_2$. 
 (We put $ \mu = 1/2$.) 
 Here $\tilde m$ denotes the order of the Pad\'e approximant. The results from the sixth- and ninth-order show good convergence up to a small fraction of a percent.}

\begin{tabular}{lllll}
\hline
\text{P1} & \text{P2} & \text{WKBPade6($\tilde m=4$)} &
   \text{WKBPade9($\tilde m=4$)} & \text{Diff,$\%$} \\
   \hline
   \hline
 \text{      1} & \text{      1} & \text{ 0.241502}-\text{ 0.063428} i & \text{
    0.241499}-\text{ 0.063425} i & 0.002 \\
  \text{      1} & \text{      2} & \text{ 0.187119}-\text{ 0.047103} i & \text{
    0.187116}-\text{ 0.047101} i & 0.002 \\
  \text{      1} & \text{      3} & \text{ 0.154893}-\text{ 0.037812} i & \text{
    0.154891}-\text{ 0.037811} i & 0.001 \\
  \text{      1} & \text{      4} & \text{ 0.133143}-\text{ 0.031705} i & \text{
    0.133142}-\text{ 0.031701} i & 0.003 \\
  \text{      1} & \text{      5} & \text{ 0.117297}-\text{ 0.02734} i & \text{
    0.117296}-\text{ 0.027339} i & 0.001 \\
  \text{      1} & \text{      6} & \text{ 0.105153}-\text{ 0.024052} i & \text{
    0.105151}-\text{ 0.024052} i & 0.002 \\
  \text{      1} & \text{      7} & \text{ 0.095501}-\text{ 0.02148} i & \text{ 
    0.0955}-\text{ 0.021478} i & 0.002 \\
  \text{      1} & \text{      8} & \text{ 0.087618}-\text{ 0.019405} i & \text{
    0.087617}-\text{ 0.019405} i & 0.001 \\
  \text{      1} & \text{      9} & \text{ 0.081042}-\text{ 0.017696} i & \text{
    0.081041}-\text{ 0.017696} i & 0.001 \\
  \text{      1} & \text{     10} & \text{ 0.07546}-\text{ 0.016263} i & \text{
    0.07546}-\text{ 0.016262} i & 0.001 \\
  \text{      5} & \text{      1} & \text{ 0.117297}-\text{ 0.02734} i & \text{
    0.117296}-\text{ 0.027339} i & 0.001 \\
  \text{      5} & \text{      2} & \text{ 0.097346}-\text{ 0.023083} i & \text{
    0.097345}-\text{ 0.023083} i & 0.001 \\
  \text{      5} & \text{      3} & \text{ 0.084573}-\text{ 0.020193} i & \text{
    0.084572}-\text{ 0.020193} i & 0.001 \\
  \text{      5} & \text{      4} & \text{ 0.075429}-\text{ 0.018042} i & \text{
    0.075428}-\text{ 0.018042} i & 0.001 \\
  \text{      5} & \text{      5} & \text{ 0.068446}-\text{ 0.016359} i & \text{
    0.068446}-\text{ 0.016359} i & 0.000 \\
  \text{      5} & \text{      6} & \text{ 0.062883}-\text{ 0.014993} i & \text{
    0.062883}-\text{ 0.014995} i & 0.003 \\
  \text{      5} & \text{      7} & \text{ 0.058314}-\text{ 0.013859} i & \text{
    0.058314}-\text{ 0.013859} i & 0.000 \\
  \text{      5} & \text{      8} & \text{ 0.054476}-\text{ 0.012898} i & \text{
    0.054476}-\text{ 0.012898} i & 0.000 \\
  \text{      5} & \text{      9} & \text{ 0.051194}-\text{ 0.012071} i & \text{
    0.051194}-\text{ 0.012071} i & 0.000 \\
  \text{      5} & \text{     10} & \text{ 0.048347}-\text{ 0.011351} i & \text{
    0.048347}-\text{ 0.011351} i & 0.000 \\
  \text{     10} & \text{      1} & \text{ 0.07546}-\text{ 0.016263} i & \text{
    0.07546}-\text{ 0.016262} i & 0.001 \\
  \text{     10} & \text{      2} & \text{ 0.064811}-\text{ 0.014558} i & \text{
    0.064811}-\text{ 0.014558} i & 0.000 \\
  \text{     10} & \text{      3} & \text{ 0.057707}-\text{ 0.013257} i & \text{
    0.057707}-\text{ 0.013257} i & 0.000 \\
  \text{     10} & \text{      4} & \text{ 0.052458}-\text{ 0.012213} i & \text{
    0.052458}-\text{ 0.012213} i & 0.000 \\
  \text{     10} & \text{      5} & \text{ 0.048347}-\text{ 0.011351} i & \text{
    0.048347}-\text{ 0.011351} i & 0.000 \\
  \text{     10} & \text{      6} & \text{ 0.045003}-\text{ 0.01062} i & \text{
    0.045003}-\text{ 0.01062} i & 0.000 \\
  \text{     10} & \text{      7} & \text{ 0.042206}-\text{ 0.009991} i & \text{
    0.042206}-\text{ 0.009992} i & 0.002 \\
  \text{     10} & \text{      8} & \text{ 0.03982}-\text{ 0.009443} i & \text{
    0.03982}-\text{ 0.009443} i & 0.000 \\
  \text{     10} & \text{      9} & \text{ 0.037751}-\text{ 0.008959} i & \text{
    0.037751}-\text{ 0.00896} i & 0.003 \\
  \text{     10} & \text{     10} & \text{ 0.035934}-\text{ 0.008529} i & \text{
    0.035934}-\text{ 0.008529} i & 0.000 \\
\hline
\hline
\end{tabular}
\label{tableQNM1}
\end{table}

\subsection{Eikonal approximation}

In the subsequent analysis, we adopt the eikonal approximation, which holds for $ l \gg 1 $. 
The extremum of the eikonal part of the effective potential is determined by condition 

\begin{eqnarray} \label{QNM6V}
\mathcal{V}'= f \frac{d \mathcal{V}}{dR}  = 0, \end{eqnarray}
or, equivalently,
\color{black}
\begin{eqnarray}           
  R^3  - 3 \mu  R^2  \nonumber \\
    + [ \mu ( -  P_1 -  P_2)  - P_1 P_2 ] R   +  \mu  P_1 P_2  = 0. 
           \label{4.masteqR}
 \end{eqnarray} 

This relation is equivalent to relation $\frac{d}{dR}\left( \frac{f}{C} \right) = 0$. It was 
proved earlier in Ref. \cite{IKMN} that
For any values of $\mu > 0$, $P_1 > 0$, and  $P_2 > 0$, the third-order 
polynomial master equation (\ref{4.masteqR}) admits one and only 
one real root $R_0$ that satisfies the inequality $R_0 > 2 \mu$. 
This result can be readily verified by expressing the master equation 
(\ref{4.masteqR}) in terms of dimensionless parameters
\begin{equation} 
  \label{4.xp}
  x = R/(2 \mu), \qquad p_i = P_i/(2 \mu),
\end{equation}
$i = 1,2$. By doing this one get \cite{IKMN}
  \begin{eqnarray}   \nonumber     
   P(x) = x^3 - \frac{3}{2} x^2 
   \qquad \qquad \qquad \qquad  \\ 
   + \left(  -  \frac{1}{2} p_1 - \frac{1}{2} p_2 -  p_1 p_2 \right) x   +  \frac{1}{2}  p_1 p_2  = 0.
   \qquad   \label{4.masteqx}
 \end{eqnarray}  

Reference \cite{IKMN} proved that, for all $p_1 > 0$ and $p_2 > 0$, 
the reduced master equation (\ref{4.masteqx}) possesses one and only one real 
root $x_{0} = R_0/(2 \mu)$ satisfying $x_{0} > 1$.

An explicit relation for   $x_{0}$  reads \cite{IKMN}
\begin{equation} 
x_0  = Y^{1/3} +  S Y^{-1/3} + \frac{1}{2},
\label{4.x3R}
\end{equation}
where
\begin{equation}
Y = i \sqrt{ {\cal R}}/(8 \times  3^{3/2}) + Z.
\label{4.Y}
\end{equation}
Here 
\begin{eqnarray} 
  {\cal R} =  64 p_1^3 p_2^3+ 96 p_1^3 p_2^2 + 96 p_1^2 p_2^3  
   \nonumber  \\
   + 48 p_1^3 p_2 + 240 p_1^2 p_2^2 + 48 p_1 p_2^3
   \nonumber  \\ 
   +8 p_1^3 + 168 p_1^2 p_2 + 168 p_1 p_2^2 + 8 p_2^3
       \nonumber  \\    
    +9 p_1^2 + 126 p_1 p_2 + 9 p_2^2,  
           \label{4.R}
 \end{eqnarray} 
and 
\begin{eqnarray} 
Z = (p_1+ p_2 +1)/8 > 1/8   \label{4.Z},\\
S = (4 p_1 p_2 +2 p_1 + 2 p_2 +3)/12 > 1/4.  \label{4.S}
 \end{eqnarray} 

By using  polar decomposition
 \begin{equation}
Y =  |Y| \exp{(i \alpha)}  
 \label{4.Y1}
 \end{equation}
with
\begin{eqnarray}
|Y| =  \sqrt{\frac{{\cal R}}{64 \times 27} + Z^2},  \label{4.modY} \\ 
\alpha = \arctan{\left(\frac{\sqrt{{\cal R}}}{8 Z \times 3^{3/2}}\right)},
\label{4.alpha}   
\end{eqnarray}
$0 < \alpha < \pi/2$, and the  identity $|Y|^{1/3} = S|Y|^{-1/3}$, or               
\begin{equation}
|Y|^2 = S^3,  \label{4.modYS}                  
\end{equation}
\color{black}
one can rewrite (\ref{4.x3R}) as follows \cite{IKMN}
\begin{equation}
x_0  =  2 |Y|^{1/3} \cos{(\alpha/3)} + \frac{1}{2}.
\label{4.x3cos}  
\end{equation}

We note that relation (\ref{4.modYS}) may be written
as \footnote[2]{Here we have eliminated the typo in formula (71) from Ref. \cite{IKMN}. }
\begin{equation}
\frac{{\cal R}}{64 \times 27} + Z^2=  S^3. 
\label{4.RS}
\end{equation}

The maximum of the eikonal part  of the effective potential thus becomes
\begin{eqnarray}
\nonumber
\mathcal{V}_0=\mathcal{V}(R_0)=
\frac{l (l+1) }{R_0^2}\left(1-\frac{2 \mu }{R_0}\right) \times \\
\times \left(1+\frac{P_1}{R_0}\right)^{-2}
\left(1+\frac{P_2}{R_0}\right)^{-2}. \label{QNM9V}
\end{eqnarray}
Here $R_0 =  2 \mu x_0$, where $x_0$ is given by (\ref{4.x3cos}).  

The second derivative with respect to the tortoise coordinate 
 (in the point of extremum) is given by
\begin{eqnarray}
\mathcal{V}_0''&=&\frac{d^2\mathcal{V}}{dR_*^2}\bigg|_{R_{*}=R_{*}(R_0)}=
f^2 \frac{d^2\mathcal{V}}{dR^2}\bigg|_{R=R_0}.
 \label{QNM10V2}
\end{eqnarray}

The second derivative $\frac{d^2\mathcal{V}}{dR^2}|_{R=R_0}$ was calculated in fact in
Ref. \cite{IKMN}
\begin{equation}
 \frac{d^2\mathcal{V}}{dR^2}\bigg|_{R=R_0} =
   - \frac{l(l+1)}{(2 \mu)^4} \frac{(6x_0^2- 6x_0 -2 p_1 p_2 - p_1  - p_2) }{(x_0+p_1)^3 (x_0+ p_2)^3}.
       \label{4.d2UdR2}
  \end{equation}
 It was shown in Ref. \cite{IKMN} that $\frac{d^2\mathcal{V}}{dR^2}|_{R=R_0} < 0$ for all $P_1 > 0$, 
 $P_2 > 0$ and $\mu >0$.
 
Thus 
\begin{equation}
\mathcal{V}_0''= - f_0^2  
\frac{l(l+1)}{(2 \mu)^4} \frac{(6x_0^2- 6x_0 -2 p_1 p_2 - p_1  - p_2) }{(x_0+p_1)^3 (x_0+ p_2)^3} <0,
 \label{QNM13}
\end{equation}
where 
\begin{eqnarray}
\nonumber
 f_0 = \left(1-\frac{2 \mu }{R_0}\right) \left(1+\frac{P_1}{R_0}\right)^{-1}
 \left(1+\frac{P_2}{R_0}\right)^{-1} \\
  = \left(1-\frac{1 }{x_0}\right) \left(1+\frac{p_1}{x_0}\right)^{-1}
  \left(1+\frac{p_2}{x_0}\right)^{-1}.
   \label{QNM14}
\end{eqnarray}

In the eikonal limit ($l \gg 1, n$), the squared frequency satisfies  \cite{2009CQGra..26p3001B,2011RvMP...83..793K}
\begin{equation}
\omega^{2}=\mathcal{V}_0-i \left(n+\frac{1}{2}\right) \sqrt{-2 \mathcal{V}_0''} + O(1),
\label{QNM12}
\end{equation}
with $n=0,1,\dots$ being the overtone number. 
Properly choosing the sign of $\omega$ yields the asymptotic ($l\to\infty$) real and imaginary parts

\begin{eqnarray}
  {\rm Re}(\omega)  &=&   \left(l + \frac{1}{2}\right) \sqrt{\nu_0}
      + O(1/l),  \label{QNM11Re} 
      \\
 {\rm Im}(\omega)  &=&  - \left(n + \frac{1}{2}\right) \frac{\sqrt{- 2 \mathcal{V}_0''}}{\sqrt{\mathcal{V}_0}}
      + O(1/l), \label{QNM11Im}
\end{eqnarray}
where $\nu_0 = \mathcal{V}_0/(l(l+1))$.

\subsection{Two limiting cases of the eikonal spectrum }\label{limitingcase2}

We consider two limits: $P_1=P_2=0$ (Schwarzschild) and $P_1=P_2=P>0$ (RN). We put $G = 1$.
In the first, the eikonal QNM relations (\ref{QNM11Re},\ref{QNM11Im}) yield 
\begin{eqnarray}\nonumber
 {\rm Re}(\omega)&=&\left(l+\frac{1}{2}\right)\sqrt{\frac{M}{r_0^3}}+O(1/l),\\ \nonumber
{\rm Im}(\omega)&=&-\left(n+\frac{1}{2}\right)\sqrt{\frac{M}{r_0^3}}+O(1/l),
\end{eqnarray} 
with $r_0=3M$ (photon sphere), matching Ref. ~\cite{1984PhLA..100..231B}. In the second, they give 
\begin{eqnarray} \nonumber
{\rm Re}(\omega)&=&\left(l+\frac{1}{2}\right)\sqrt{\frac{M}{r_0^3}-\frac{Q^2}{2 r_0^4}}+ O(1/l),\\
{\rm Im}(\omega)&=&-\left(n+\frac{1}{2}\right)\sqrt{\frac{M}{r_0^3}-\frac{Q^2}{2 r_0^4}} \sqrt{\frac{3 M}{r_0}-\frac{2 Q^2}{r_0^2}} + O(1/l) \nonumber
\end{eqnarray}
with $r_0=\frac{3M}{2}+\frac12\sqrt{9M^2-4Q^2}=R_0+P$ (photon orbit), 
compatible with Ref. ~\cite{1996PhRvD..54.7470A} ($n=0$) upon $Q^2=2Q_{RN}^2$.

\section{Conclusions}\label{conclusion}

We have examined a non-extremal black hole dyon-like solution in a 4-dimensional gravitational model 
with two scalar fields and two Abelian vector fields which was considered (in composite version) 
in Ref. ~\cite{IKMN}. The model contains two vectors of dilatonic coupling vectors 
$\vec{\lambda}_s $, $s =1,2$, obeying $\vec{\lambda}_1 = \vec{\lambda}_2 = \vec{\lambda}$ and 
$\vec{\lambda}^2 = \frac{1}{2}$. 
In fact this is a solution with independent electric and magnetic charges.

We have also presented several physical parameters of the solution, namely the gravitational mass $M$, the scalar charges $Q_{\varphi}^i$, the Hawking temperature, and the black hole area entropy. In addition, we have considered the first law of black hole thermodynamics and verified the validity of the Smarr relation for our solution.

We have investigated the solutions of the massless Klein–Fock–Gordon equation for the test scalar field in the background of  the metric of our static, spherically symmetric solution. By using the method of separation of variables and the tortoise radial coordinate the Klein–Fock–Gordon equation is simplified, leading to a radial equation governed by an effective potential. This potential depends on the solution parameters, such as 
$P_1 >0$, $P_2 >0$, $\mu >0$.
The physical quantities—including mass, electric and magnetic (color) charges, scalar charges -- are determined by some of these parameters. 

Here, we focused mainly on the eikonal part of the effective potential and employed the value of the radial coordinate (radius) 
$R_0$   corresponding to the maximum of this part, which was previously found in Ref. ~\cite{IKMN}.

Using the maximum of the effective potential in the eikonal regime, its corresponding radius and the second derivative 
of the eikonal part of the effective potential in the point of maximum ~\cite{IKMN}, we have calculated the cyclic frequencies of the quasinormal modes within the eikonal approximation. We have also explored two limiting cases that reduce to the Schwarzschild and Reissner-Nordström solutions, corresponding to when the solution parameters assume the two distinct values:  $P_1 = P_2= +0$  and  $P_1 = P_2 = P > 0$,
 respectively. As a result, we have ascertained that our outcomes on eikonal spectrum are in full agreement with the earlier findings in the literature.

 Here we have also presented the results of numerical calculations of QNM frequencies by using higher-order WKB formula 
 ~\cite{2019CQGra..36o5002K} for two lower levels $n=0$, $\ell=0$, and $n=0$, $\ell=1$,  for various $P_1$, $P_2$.  
 The reliability of the approximations used is established by comparing the results from two different orders of the WKB method, which demonstrate excellent convergence especially for the case $\ell>n$ (up to a small fraction of a percent).

 \begin{acknowledgements}
 
 We thank Prof. K. Bronnikov for valuable comments.
 
 \end{acknowledgements}

 \end{document}